\documentclass[aps,prd,showpacs,amsmath,amssymb,onecolumn,10pt]{revtex4}
\usepackage{graphicx}
\usepackage{amsmath}
\usepackage{amssymb}
\usepackage{amsfonts}
\usepackage{bm}
\usepackage{color}
\usepackage[colorlinks, linkcolor={blue},citecolor={blue}]{hyperref}

\def\be{\begin{equation}}
\def\ee{\end{equation}}
\def\bea{\begin{eqnarray}}
\def\eea{\end{eqnarray}}

\newcommand{\f}[2]{\frac{#1}{#2}}

\begin{document}

\title{The Weyl-Cartan Gauss-Bonnet gravity}
\author{Zahra Haghani$^1$}
\email{z.haghani@du.ac.ir}
\author{Nima Khosravi$^2$}
\email{nima@ipm.ir}
\author{Shahab Shahidi$^1$}
\email{s.shahidi@du.ac.ir}
\affiliation{$^1$ School of Physics, Damghan University, Damghan, 
41167-36716, Iran}
\affiliation{$^2$ School of Astronomy, Institute for Research in Fundamental
Sciences (IPM), P.O. Box 19395-5531, Tehran, Iran}
\date{\today}
\begin{abstract}
In this paper, we consider the generalized Gauss-Bonnet action in 
$4$-dimensional Weyl-Cartan space-time. In this space-time, the presence of 
torsion tensor and Weyl vector implies that the generalized 
Gauss-Bonnet action will not be a total derivative in four dimension 
space-time.
It will be shown that the higher than two time derivatives can be removed from 
the 
action by choosing suitable set of parameters. In the special case where
only the trace part of the torsion remains, the model reduces to GR plus two 
vector fields. One of which is
massless and the other is  massive. We will then obtain the healthy region of 
the 5-dimensional parameter space of the theory in some special cases.
\end{abstract}
\maketitle

\section{Introduction}
In 1918 Weyl proposed a new geometry to unify electromagnetism with Einstein's 
general relativity \cite{weyl}. 
In Riemannian geometry one has a priori condition that the length of a vector 
should not change during the parallel transportation.
In the Weyl geometry, this assumption is dropped and so a parallel 
transported 
vector has different length and direction with respect to the original vector. 
The gravitational theory which is
built on the Weyl geometry is known as the Einstein-Weyl gravity \cite{weyl}. 
In Einstein-Weyl gravity the connection is no longer metric compatible, so, the 
covariant derivative of the metric is not zero. Instead one has the 
relation
\begin{align}
\tilde{\nabla}_\mu g_{\nu\rho}=Q_{\mu\nu\rho},
\end{align}
where the tensor $Q_{\mu\nu\rho}$ is symmetric with respect to its last two 
indices. Weyl proposed the special case $Q_{\mu\nu\rho}\propto w_\mu 
g_{\nu\rho}$ for his theory where $w_\mu$ is the Weyl vector. One of the 
important consequences of this 
geometry is that the unit vector changes through parallel transportation.
Suppose that the length of an arbitrary vector field $A^\mu$ is $l$. 
During the parallel transportation, the variation of the length of $A^\mu$ can 
be written in terms of the Weyl vector as
\begin{align}
dl=lw_\mu dx^\mu.
\end{align}
For a closed curve, the length of the vector $A^\mu$ changes 
as
\begin{align}
l\rightarrow l-\int_S lW_{\mu\nu}dS^{\mu\nu},
\end{align}
where $S$ is the area of the closed curve, $dS^{\mu\nu}$ is the infinitesimal 
element of the surface, and
\begin{align}
W_{\mu\nu}=\partial_\mu w_\nu-\partial_\nu w_\mu,
\end{align}
is called the Weyl's length curvature which is the same as the electromagnetic 
field strength. This implies that one has the freedom to 
choose the unit length at each point, which is the Weyl gauge freedom 
\cite{weyl}. A variety of works have been done in the Weyl geometry including 
the cosmology \cite{ww1}, relations to scalar-tensor \cite{ww2} and 
teleparallel theories \cite{ww3}.

One can also restrict the form of Weyl vector to be a derivative of a scalar as 
$w_\mu=\partial_\mu\phi$ \cite{weylint}. In this case the length curvature 
$W_{\mu\nu}$ vanishes and one can then 
define a fixed unit length at each point. We note that the unit length varies 
at 
different points. The resulting theory is known as the Weyl 
integrable theory \cite{wi1}.

Another generalization of Einstein gravity can be proposed by assuming the
existence of an asymmetric connection on the space-time manifold. The first 
attempt for this purpose is due to Eddington in 
1921 in order to generalize the Einstein's general relativity to get some 
insights about microscopic physics \cite{edi}. The major attempt in this way 
was done by 
Cartan in 1922 where he defined the torsion tensor as the antisymmetric part of 
the general connection \cite{cartan}. The theory based on this assumption is 
called the 
Einstein-Cartan theory. Cartan believed that the torsion tensor should be 
related in some way to the intrinsic angular momentum of the matter content of 
the 
universe. So, the torsion should vanish in the absence of matter 
\cite{cartan}. However, this idea had been forgotten for  a 
while.  The concept of the asymmetric affine connection came back to the 
literature as a way to write a unified field theory of  general relativistic 
type \cite{unified}. 
The energy momentum tensor of a spinning massive particle was known to be 
asymmetric \cite{costa,weys}.  
It was clear that  these
particles can not be a source of Einstein's general relativity equation. 
In order to consider these particles in the context of general relativity, one 
should generalize the Einstein's equation to a theory with asymmetric 
energy-momentum tensor. The concept of torsion, can then be used as a way to 
consider these type of particles \cite{masi}. Many works have been done in the 
context of theories with torsion, including teleparallel theories \cite{tele}, 
and Weyl-Cartan theories \cite{zahra, isra}. For a 
very good review on the subject of Cartan theory, see \cite{hehl}.

There is another way to generalize the Einstein's general relativity, by 
replacing the Ricci scalar with a general function of it, known as  $f(R)$ 
gravity theories \cite{fR,fR1}. But one can add to the action, any scalar 
combination of the Riemann tensor and its contractions. One of the first 
attempts 
was 
done by Kretschmann \cite{krech} in 1917 by introducing the action of the form
\begin{align}\label{}
S_K=\int d^4x \sqrt{-g} R_{\mu\nu\rho\sigma}R^{\mu\nu\rho\sigma},
\end{align}
instead of the Einstein-Hilbert action. The above action has higher than second 
order time derivatives of the metric in its field equation and hence contains 
ghost instabilities. It turns out that the unique combination of two Riemann
tensors and its contractions which leads to at most second order time 
derivatives in the field equation is of the form
\begin{align}\label{}
S_{GB}=\int d^nx \sqrt{-g} 
(R_{\mu\nu\rho\sigma}R^{\mu\nu\rho\sigma}-4R_{\mu\nu}R^{\mu\nu}+R^2),
\end{align}
which is called the Gauss-Bonnet term. In four space-time dimensions, this term 
can be 
written as a total derivative, and can be dropped from the field equations 
\cite{fR1}. This leads to the conclusion that in 4D, the Riemannian geometry 
together with the condition of stability has a unique candidate for the 
gravitational action, which is the Einstein-Hilbert action.

In non-Riemannian geometries such as Weyl and Cartan geometries, the above 
conclusion is no longer true and the Gauss-Bonnet term will not become total 
derivative. In \cite{tomi} the Gauss-Bonnet combination was obtained in the 
context of Weyl geometry, and it turns out that the remaining term in 4D is the 
Weyl vector kinetic term
\begin{align}\label{}
S_{GB}\propto \int d^4x \sqrt{-g} W_{\mu\nu}W^{\mu\nu}.
\end{align}
It is the aim of the present paper to generalize the above argument to the case 
of Weyl-Cartan space-time. Similar to the Einstein-Weyl space-time, in the 
Weyl-Cartan space-time the Gauss-Bonnet term will not be a total derivative. 
The 
theory is not in general Ostrogradski stable. In order to have a stable theory 
one should constrain the parameter space of the model as we will do in the 
following
sections.

The Weyl-Cartan model has also been considered in the context of Weitzenboch 
gravity in \cite{zahra}. The authors have added the kinetic terms for 
the Weyl vector and the torsion tensor by hand, using the trace of torsion 
tensor.  The main purpose of the present work is to see if adding higher order 
gravity terms in 
Lagrangian can show a way to have kinetic terms automatically. For this case we 
will just add the 
Gauss-Bonnet term which we know does not produce any ghost for the metric 
perturbations. Adding non-metricity 
as well as torsion has been
studied in \cite{ref1}. The author has shown that at the second order 
gravity level,  connection has no dynamics which is same as the results in 
\cite{zahra}. Higher order curvature terms in 
the presence of torsion (without non-metricity) has been studied in 
\cite{ref2} where the authors have shown that chiral fermionic 
matter fields can live in a Riemann-Cartan geometry. Particularly, the Lovelock 
theory in the presence of the torsion has been studied in the $d$-dimensions 
\cite{ref3} and in the context of stringy fluxes \cite{ref4}. 

Before going ahead 
let us briefly say our approach. In general, we are working in the context of 
effective field theory. However there are two (equivalent) approaches which are 
extensively studied, for example,in the context of inflation. In the first 
approach \cite{senatore} all the possible perturbations have been assumed at the 
level of the Lagrangian with arbitrary coefficients. The equivalent alternative  
approach is to assume the possible terms at the level of background 
\cite{weinberg}. Here we will do more or less the same approach as in  
\cite{weinberg}. We first assume we have just a metric and try to have 
all the consistent possible terms at the level of the Lagrangian which is Ricci 
scalar and Gauss-Bonnet (in general Lovelock terms). Then we ask what happens 
to this model if one assume non-metricity and torsion. The other approach, 
which is more similar to \cite{senatore}, is to write all the possible terms 
constructed by non-metricity and torsion in addition to the metric itself. This 
second approach was done in \cite{ref1,ref2}.

We will 
see in this paper that considering the Gauss-Bonnet action can produce 
automatically all the 
kinetic terms of \cite{zahra}. It is worth mentioning that the theory 
\cite{zahra} has a potential ghost, noting that in the Weitzenboch gravity the 
torsion has some relation to the Ricci scalar, and as a result the torsion 
kinetic term has more than second time derivatives. However, the present paper 
is free from the aforementioned instability due to the absence of 
the Weitzenboch condition.
One should note that the torsion self-interaction term $\nabla_\mu T \nabla^\mu 
T$ with $T=T^\mu T_\mu$ and $T_\mu=T^\nu_{~\mu\nu}$, in \cite{zahra} 
can not be produced in the present context, because it is fourth order in the 
torsion and second order in derivatives. In order to produce such term one 
should consider the higher order Lovelock terms in the action.

The present theory in general may have some tachyon instabilities but the 
analysis is very complicated because of the appearance of the torsion tensor. 
In section \ref{sec1} the generalized Gauss-Bonnet  action in 
Weyl-Cartan space-time is introduced and shown that the higher than two  
time 
derivatives are removed in the action. 
In section \ref{sec2} we will consider a restricted form for the torsion tensor 
and obtain the healthy region of the parameter space in which tachyonic 
instabilities 
are removed.
\section{The model}\label{sec1}
The Weyl geometry proposal induces a new vector which results in non-metricity 
of the 
connection i.e. $\bar{\nabla}_\mu g_{\alpha\beta}\neq 0$ where 
$\bar{\nabla}_\mu$ is the 
covariant derivative with respect to the Weyl connection. 
Mathematically, the Weyl geometry has a special form of non-metricity i.e.
\begin{align}\label{mo1}
\bar{\nabla}_\mu g_{\nu\sigma}=2w_\mu g_{\nu\sigma},
\end{align}
where $w_{\mu}$ is the Weyl vector. So the Weyl connection can be obtained as
\begin{align}\label{mo2}
\bar{\Gamma}^\lambda_{~\mu\nu}=\left\{\begin{smallmatrix}\lambda\\ 
\mu~\nu\end{smallmatrix}\right\}
+Q^\lambda_{\mu\nu},
\end{align}
where
\begin{align}\label{mo2.1}
Q^\lambda_{\mu\nu}=g_{\mu\nu}w^\lambda-\delta^\lambda_\mu 
w_\nu-\delta^\lambda_\nu 
w_\mu,
\end{align}
and $\left\{\begin{smallmatrix}\lambda\\ 
\mu~\nu\end{smallmatrix}\right\}$ is the Christoffel symbol. In addition one 
may generalize the above 
connection by adding the effects of the torsion into it as
\begin{align}\label{mo2}
\Gamma^\lambda_{~\mu\nu}=\left\{\begin{smallmatrix}\lambda\\ 
\mu~\nu\end{smallmatrix}\right\}
+Q^\lambda_{\mu\nu}+C^\lambda_{\mu\nu}.
\end{align}
Note that the third term is named contortion tensor defined as
\begin{align}\label{mo3}
C^\lambda_{~\mu\nu}=T^\lambda_{~\mu\nu}-g^{\lambda\beta}g_{\sigma\mu}T^\sigma_{
~\beta\nu}-g^{\lambda\beta}g_{\sigma\nu}T^\sigma_{~\beta\mu},
\end{align}
where we have defined the torsion tensor $T^\lambda_{~\mu\nu}$ as
\begin{align}\label{mo4}
T^\lambda_{~\mu\nu}=\f{1}{2}\left(\Gamma^\lambda_{~\mu\nu}-\Gamma^\lambda_{
~\nu\mu}\right).
\end{align}
It is easy to show that the additional torsion does not affect the 
non-metricity 
relation i.e. the relation (\ref{mo1}) is still valid. 
By using the metric one can build 
$C_{\lambda\mu\nu}=g_{\lambda\sigma}C^\sigma_{~\mu\nu}$ which is antisymmetric 
with respect to its two first indices by having in mind that the torsion tensor 
is antisymmetric with respect to its lower indices in $T^\sigma_{~\mu\nu}$.

We define the curvature tensor as
\begin{align}\label{mo5}
K^\lambda_{~\mu\nu\sigma}=\partial_\nu\Gamma^\lambda_{~\mu\sigma}
-\partial_\sigma\Gamma^\lambda_{~\mu\nu}
+\Gamma^\alpha_{~\mu\sigma}\Gamma^\lambda_{~\alpha\nu}-\Gamma^\alpha_{~\mu\nu}
\Gamma^\lambda_{~\alpha\sigma}.
\end{align}
One can decompose the curvature tensor into four parts as
\begin{align}\label{mo6}
K^\lambda_{~\mu\nu\sigma}=R^\lambda_{~\mu\nu\sigma}+C^\lambda_{~\mu\nu\sigma}
+Q^\lambda_{~\mu\nu\sigma}+I^\lambda_{~\mu\nu\sigma},
\end{align}
where the first term in the right hand side of the above relation is the 
Riemann 
curvature tensor defined by the 
Christoffel symbol and we have defined
\begin{align}
C^\lambda_{~\mu\nu\sigma}&=\nabla_\nu C ^\lambda_{~\mu\sigma}-\nabla_\sigma C 
^\lambda_{~\mu\nu}
+ C ^\alpha_{~\mu\sigma} C ^\lambda_{~\alpha\nu}- C ^\alpha_{~\mu\nu} C 
^\lambda_{~\alpha\sigma},\label{mo7}\\
Q^\lambda_{~\mu\nu\sigma}&=\nabla_\nu Q ^\lambda_{~\mu\sigma}-\nabla_\sigma Q 
^\lambda_{~\mu\nu}
+ Q ^\alpha_{~\mu\sigma} Q ^\lambda_{~\alpha\nu}- Q ^\alpha_{~\mu\nu} Q 
^\lambda_{~\alpha\sigma},\label{mo8}\\
I^\lambda_{~\mu\nu\sigma}&=C^\alpha_{~\mu\sigma}Q^\lambda_{~\alpha\nu}+Q^\alpha_
{~\mu\sigma}C^\lambda_{~\alpha\nu}
-C^\alpha_{~\mu\nu}Q^\lambda_{~\alpha\sigma}-Q^\alpha_{~\mu\nu}C^\lambda_{
~\alpha\sigma},\label{mo9}
\end{align}
where $\nabla_\mu$ is the covariant derivative with respect to the Christoffel 
symbol and $I^\lambda_{~\mu\nu\sigma}$ represents interaction between 
non-metricity and torsion parts. It is possible to rewrite the purely 
non-metricity part  \eqref{mo8} as
\begin{align}\label{mo10}
\f{1}{2}Q^\lambda_{~\mu\nu\sigma}=-\delta^\lambda_\mu\nabla_{[\nu}w_{\sigma]}
-\delta^\lambda_{[\sigma}\nabla_{\nu]}w_\mu
-g_{\mu[\nu}\nabla_{\sigma]}w^\lambda+\delta^\lambda_{[\nu}w_{\sigma]}w_\mu+g_{\mu[\nu}\delta^\lambda_{\sigma]}w^2
+g_{\mu[\sigma}w_{\nu]}w^\lambda,
\end{align}
where $w^2=w_\mu w^\mu$ and the interaction part  \eqref{mo9} as
\begin{align}\label{mo11}
\f{1}{2}I^\lambda_{~\mu\nu\sigma}=-w^\alpha 
C_{\alpha\mu[\sigma}\delta^\lambda_{\nu]}
-w^\alpha C^\lambda_{~\alpha[\sigma}g_{\nu]\mu}-w^\lambda 
C_{\mu[\nu\sigma]}-w_\mu C^\lambda_{~[\sigma\nu]}.
\end{align}
Note that the 
curvature tensor $K^\mu_{~\nu\rho\sigma}$ is still antisymmetric wrt its last 
two indices. 

In order to construct a higher order gravity models e.g. Gauss-Bonnet action, 
one should multiply the curvature tensor to itself. There are seven 
different ways to do this
\begin{align}\label{mo12}
K_{\lambda\mu\nu\sigma}K^{\lambda\mu\nu\sigma},K_{\lambda\mu\nu\sigma}K^{
\mu\lambda\nu\sigma},K_{\lambda\mu\nu\sigma}K^{\nu\sigma\lambda\mu},K_{\lambda\mu\nu\sigma}K^{\lambda\nu\mu\sigma},K_{
\lambda\mu\nu\sigma}K^{\nu\mu\lambda\sigma},K_{\lambda\mu\nu\sigma}K^{
\mu\sigma\lambda\nu},K_{\lambda\mu\nu\sigma}K^{\sigma\lambda\mu\nu}.
\end{align}
One should note that in the case of vanishing Weyl and torsion, only the first three terms of equation \eqref{mo12} can be reduced to $K_{\lambda\mu\nu\sigma}K^{\lambda\mu\nu\sigma}$ which is present in the standard Gauss-Bonnet Lagrangian. Adding the rest of these combinations to the Lagrangian will produce higher than second order time derivatives of the metric which makes our theory unstable. So, we will ignore them in the following and only add the first three terms to the action.

Now consider the contractions of different parts of the curvature tensor. It is well-known that the Riemann tensor has only one independent contraction. 
The 
Weyl part of the above curvature tensor has two independent contractions
\begin{align}\label{mo13}
&Q^\lambda_{~\lambda\mu\nu}=-4W_{\mu\nu},\nonumber\\
&Q^\lambda_{~\mu\lambda\nu}=-\nabla_\mu 
w_\nu+3\nabla_\nu 
w_\mu+g_{\mu\nu}\nabla_\lambda w^\lambda+2w_\mu w_\nu-2g_{\mu\nu}w^2,
\end{align}
where we have defined
\begin{align}\label{mo14}
W_{\mu\nu}=\nabla_\mu w_\nu-\nabla_\nu w_\mu.
\end{align}
The contortion part of the curvature tensor has only one independent contraction
\begin{align}\label{mo15}
& C^\lambda_{~\lambda\mu\nu}=0,\nonumber\\
& C^\lambda_{~\mu\lambda\nu}=\nabla_\lambda 
C^\lambda_{~\mu\nu}+\nabla_\nu C_{\mu},
\end{align}
where we have defined $C^\mu=C^{\mu\nu}_{~~~\nu}$.

The interaction part has also one independent contraction which can be written 
as
\begin{align}\label{mo16}
& I^\lambda_{~\lambda\mu\nu}=0,\nonumber\\
& I^\lambda_{~\mu\lambda\nu}=-w^\alpha(C_{\alpha\mu\nu}+C_{\nu\mu\alpha}
).
\end{align}
For the Riemann curvature tensor, we have
$R^\lambda_{~\lambda\mu\nu}=0$ and  $R^\lambda_{~\mu\lambda\nu}=R_{\mu\nu}$ 
where $R_{\mu\nu}$ is the standard Ricci tensor.
For the contracted curvature tensor, the two independent contractions 
are 
$$K_{\mu\nu}\equiv K^\lambda_{~\lambda\mu\nu},\quad 
\mathcal{K}_{\mu\nu}\equiv {K}^\lambda_{~\mu\lambda\nu}.$$
There are four 
independent combinations of them as follows
\begin{align}\label{mo16.4}
K_{\mu\nu}{K}^{\mu\nu},~K_{\mu\nu}{\mathcal{K}}^{\mu\nu},~\mathcal{K}_{\mu\nu}{
\mathcal{K}} ^ {
\mu\nu},~
\mathcal{K}_{\mu\nu} {\mathcal{K}}^{\nu\mu}.
\end{align}
In the case of vanishing Weyl and torsion, the first two terms of equation \eqref{mo16.4} vanishes, and the rest becomes identical to $R_{\mu\nu}R^{\mu\nu}$.

There is only one independent curvature scalar of the tensor 
$K^\lambda_{~\mu\nu\sigma}$ which can be 
defined by contracting the tensor ${\cal{K}}_{\mu\nu}$ with the metric
\begin{align}\label{mo17}
K=R&+6\nabla_\mu w^\mu-6w^2+2\nabla_\lambda 
C^{\lambda}-C^{\alpha}C_{\alpha}
+C_{\alpha\mu\lambda}C^{\alpha\lambda\mu}
-4w^\alpha C_{\alpha}.
\end{align}

Let us propose the following action
\begin{align}
S=\f{1}{2\kappa^2}\int d^4x\sqrt{-g}K+S_G,
\end{align}
where $S_G$ is the Gauss-Bonnet action defined as
\begin{align}\label{13}
S_G=&\int d^4x\sqrt{-g}\bigg[\alpha_1 K^{\alpha \beta 
\gamma\delta} K_{\alpha \beta \gamma\delta} + \alpha_2 K^{\alpha \beta 
\gamma\delta} 
K_{\gamma \delta \alpha \beta} 
- \alpha_3 K^{\alpha \beta \gamma\delta} K_{\beta \alpha 
\gamma\delta} 
- 4 \beta_1 {\mathcal{K}}_{\beta\gamma} {\mathcal{K}}^{\beta\gamma} 
-4 
\beta_2 {\mathcal{K}}_{\beta\gamma} 
{\mathcal{K}}^{\gamma\beta}
\nonumber\\&-4 \beta_3  {K}_{\alpha \beta} {K}^{\alpha \beta}-4 
\beta_4 
{K}_{\alpha \beta} {\mathcal{K}}^{\alpha\beta}
+ K^2\bigg],
\end{align}
where $\alpha_i$ and $\beta_i$ are arbitrary constants. To get the standard Gauss-Bonnet action in the absence of torsion and 
non-metricity we need to impose the following constraints on the coefficients 
\begin{subequations}\label{14}
\begin{align}
&\alpha_1+\alpha_2+\alpha_3=1,\\
&\beta_1+\beta_2=1.
\end{align}
\end{subequations}
The coefficients $\beta_3$ and $\beta_4$ does not enter to the above 
conditions, because in the case of vanishing Weyl vector, the tensor 
$K_{\mu\nu}$ vanishes. We should note that the above action is the most general 
action for the 
second order higher gravity in the Weyl-Cartan theory which reduces to the 
standard 
Gauss-Bonnet action in the limit of zero Weyl and torsion.

The potentially dangerous terms which can produce Ostrogradski ghost can be 
written as
\begin{widetext}
\begin{align}\label{50}
S_G\supset 4\int d^4x 
\sqrt{-g}&\bigg(-2(\beta_1+\beta_2)R_{\mu\nu}\nabla^\mu 
C^\nu+R\nabla^\mu 
C_\mu-2\big[2(\beta_1+\beta_2)-(\alpha_1+\alpha_2+\alpha_3)\big]R_{\mu\nu}
\nabla^\mu 
w^\nu\nonumber\\&
+\big[3-2(\beta_1+\beta_2)\big]R\nabla^\mu 
w_\mu-2(\beta_1+\beta_2)R_{\mu\nu}\nabla^\alpha 
C_{\alpha}^{~~\mu\nu}+(\alpha_1+\alpha_2+\alpha_3)R_{\alpha\delta\beta\gamma}
\nabla^\alpha 
C^{\beta\gamma\delta}\bigg).
\end{align}
\end{widetext}
Using equations \eqref{14}, one can write the RHS of \eqref{50} as
\begin{align}\label{50.1}
-8\int d^4x 
\sqrt{-g}&\bigg(G_{\mu\nu}\nabla^\mu(w^\nu+C^\nu)+R_{\mu\nu}
\nabla^\alpha 
C_{\alpha}^{~~\mu\nu}-\f12R_{\alpha\delta\beta\gamma}
\nabla^\alpha 
C^{\beta\gamma\delta}\bigg).
\end{align}
The first term of the above action does not contribute to the theory after 
integrating by part and using the contracted Bianchi identity. Also, by using 
the first Bianchi identity and integrating by part, one can easily show that
\begin{align}\label{50.2}
\int d^4x 
\sqrt{-g}R_{\alpha\delta\beta\gamma}
\nabla^\alpha 
C^{\beta\gamma\delta}=2\int d^4x 
\sqrt{-g} R^{\sigma\lambda}\nabla^\mu C_{\mu\lambda\sigma}.
\end{align}
Using the above equation, one can see that the last two terms in \eqref{50.1} 
cancel each other. So, the potentially dangerous terms, do not contribute to 
the full action.

In order to write the action $S_G$ in detail, we decompose the action into 
three 
parts. The terms which involve only the Weyl vector can be collected as
\begin{align}\label{4}
S_{W}= \rho \int d^4x
\sqrt{-g} W_{\mu\nu}W^{\mu\nu},
\end{align}
with
$$\rho=-4(3+2\alpha_2+2\alpha_3-8\beta_2+16\beta_3+8\beta_4),$$
and we have dropped the total derivative terms.
The terms which involves the contortion tensor can be written as
\begin{widetext}
\begin{align}\label{5}
S_{C}&=\int d^4x
\sqrt{-g}\bigg[
- 4\, {R}^{\alpha  \beta \gamma \delta}  C _{\alpha }\,^{\nu}\,_{\gamma} 
C _{\beta\nu \delta} - 8\, {R}^{\alpha  \beta} C ^{\gamma}  C _{\alpha  \gamma 
\beta} + 8\, {R}^{\alpha 
\beta}  C _{\alpha }\,^{\gamma \delta}  C _{\gamma\delta\beta}  + 2\, R  C 
^{\alpha  \beta \gamma}  C _{\alpha  
\gamma \beta} + 4\, {G}^{\alpha  \beta}  C _{\alpha }  C _{\beta}
\nonumber\\&+ (2 - 2\, \alpha_2)
\bigg( \nabla ^{\alpha }{ C ^{\beta\gamma\delta}}\,  \nabla _{\alpha }{ C 
_{\beta \gamma \delta}}\,  
- \nabla ^{\alpha }{ C ^{\beta \gamma \delta}}\,  \nabla _{\delta}{ C _{\beta 
\gamma \alpha }} -4 \nabla ^{\alpha }{ C ^{\beta \gamma 
\delta}}\,   C _{\beta}\,^{\nu}\,_{\alpha }  C _{\gamma\nu \delta} +  C 
^{\alpha  \beta \gamma}  C _{\alpha }\,^{\delta}\,_{\gamma}  C _{\beta}\,^{\nu 
\mu}  C _{\delta\nu \mu}
 \nonumber\\&-  C ^{\alpha  
\beta \gamma}  C _{\alpha }\,^{\delta \nu}  C _{\beta}\,^{\mu}\,_{\nu}  C 
_{\delta \mu \gamma} \bigg)
+ 4\,\alpha_2 \nabla ^{\alpha }{ C ^{\beta \gamma \delta}}\,  
\nabla _{\beta}{ C 
_{\alpha\delta \gamma}}   - 8\,\alpha_2 \nabla ^{\alpha }{ C ^{\beta \gamma 
\delta}}\,   C _{\alpha }\,^{\nu}\,_{\beta}  C _{\delta\nu \gamma}  + 
2\, \alpha_2  C^{\alpha  \beta \gamma}  C _{\alpha }\,^{\delta \nu}  C 
_{\gamma}\,^{\mu}\,_{\beta}  C _{\nu \mu \delta}
 \nonumber\\& - 2\, \alpha_2 C ^{\alpha  \beta \gamma}  C _{\alpha }\,^{\delta 
\nu}  C _{\gamma}\,^{\mu}\,_{\delta}  C _{\nu 
\mu \beta}   + (4 - 4\, \beta_2) \bigg(
\nabla ^{\alpha }{ C _{\alpha }\,^{\beta \gamma}}\,  \nabla ^{\delta}{ C 
_{\beta 
\delta \gamma}}\, + 2 \nabla ^{\alpha }{ C ^{\beta}}\, \nabla 
^{\gamma}{ C _{\beta \gamma \alpha }}\,  -2 \nabla ^{\alpha 
}{ C _{\alpha }\,^{\beta \gamma}}\,   C ^{\delta}  C _{\beta\delta\gamma}
\nonumber\\&+ 2 \nabla ^{\alpha }{ C _{\alpha }\,^{\beta \gamma}}\,   C 
_{\beta}\,^{\delta \nu}  C _{\delta\nu \gamma} 
 -2 \nabla^{\alpha }{ C 
^{\beta}}\,   C ^{\gamma}  C _{\beta \gamma \alpha } + 2 \nabla 
^{\alpha }{ C ^{\beta}}\,   C _{\beta}\,^{\gamma \delta} C _{\gamma\delta\alpha 
} -  C ^{\alpha }  C ^{\beta}  C _{\alpha }\,^{\gamma \delta}  
C_{\beta \gamma \delta} -2  C ^{\alpha }  C 
_{\alpha 
}\,^{\beta \gamma}  C _{\beta}\,^{\delta \nu}  C _{\delta\nu \gamma}
\nonumber\\&+C ^{\alpha  \beta \gamma}  C _{\alpha  \gamma}\,^{\delta}  C 
_{\beta}\,^{\nu \mu}  C _{\nu \mu \delta} -\f{1}{2} C 
^{\alpha  \beta}  C _{\alpha  \beta} \, \bigg)
 + 4\, \beta_2 \nabla ^{\alpha }{ C 
_{\alpha 
}\,^{\beta \gamma}}\,  \nabla ^{\delta}{ C _{\gamma\delta\beta}}\,  + 
8\,  \beta_2
\nabla ^{\alpha }{ C ^{\beta}}\,  \nabla ^{\gamma}{ C _{\alpha  \gamma 
\beta}}\, 
  - 8\, \beta_2 
\nabla ^{\alpha }{ C ^{\beta}}\,  C ^{\gamma}  C _{\alpha  \gamma \beta}
\nonumber\\& - 8\, \beta_2 \nabla ^{\alpha }{ C _{\alpha }\,^{\beta \gamma}}\,  
 C ^{\delta} 
C _{\gamma\delta\beta}+ 8\,\beta_2 \nabla ^{\alpha }{ C _{\alpha 
}\,^{\beta 
\gamma}}\,   C _{\gamma}\,^{\delta \nu}  C _{\delta\nu \beta}   
+ 8\, \beta_2 \nabla ^{\alpha }{ C ^{\beta}}\,   C _{\alpha }\,^{\gamma \delta} 
 C 
_{\gamma\delta\beta} 
- 4\, \beta_2 C ^{\alpha }  C ^{\beta}  C 
_{\alpha 
}\,^{\gamma 
\delta}  C _{\beta\delta\gamma} 
\nonumber\\& - 8\, \beta_2  C ^{\alpha }  C _{\alpha 
}\,^{\beta 
\gamma}  C _{\gamma}\,^{\delta \nu}  C _{\delta \nu \beta}  
+4\,\beta_2  C 
^{\alpha  \beta \gamma}  C _{\alpha  \gamma}\,^{\delta}  C _{\delta}\,^{\nu 
\mu} 
C _{\nu \mu \beta}    - 4\,C ^2 \nabla 
^{\alpha 
}{ C _{\alpha }}
+ 4\, \nabla ^{\alpha }{ 
C _{\alpha 
}}\,   C ^{\beta \gamma \delta}  C _{\beta\delta\gamma} +  C ^4 - 2\,  C ^2  C 
^{\beta \gamma 
\delta}  C _{\beta\delta\gamma} 
\nonumber\\&+  C ^{\alpha  \beta \gamma}  C _{\alpha  
\gamma 
\beta}  C ^{\delta\nu \mu}  C _{\delta \mu \nu}
\bigg],
\end{align}
\end{widetext}
where we have defined $C^2=C_\mu C^\mu$ and
\begin{align}\label{6}
C_{\mu\nu}=\nabla_\mu C_\nu -\nabla_\nu C_\mu.
\end{align}
One should note that the tensor $C_{\mu\nu}$ is proportional to the tensor 
$T_{\mu\nu}$ constructed similarly with the torsion tensor. One can see that 
the term $T_{\mu\nu}T^{\mu\nu}$ is produced naturally in this model, which is 
also the kinetic term assumed in \cite{zahra}.

The remaining terms of the action contain a variety of possible interactions 
between 
the Weyl vector and the contortion tensor
\begin{widetext}
\begin{align}\label{a1}
S_{I}=&\int 
d^4x\sqrt{-g}\bigg[
- 8\, {R}^{\alpha \beta\gamma\delta}  C _{\alpha  \gamma\beta}  w _{\delta} + 
(  8\, \beta_2 - 8\, \alpha_2) \big({\nabla}^{\alpha }{ C _{\alpha 
}\,^{\beta\gamma}}\,   C _{\beta}\,^{\delta}\,_{\gamma} 
 w _{\delta} - C 
^{\alpha \beta}\,   C 
_{\alpha}\,^{\gamma}\,_{\beta }  w _{\gamma} -  C ^{\alpha }  C _{\alpha 
}\,^{\beta\gamma}  C 
_{\beta}\,^{\delta}\,_{\gamma}  w _{\delta}
\nonumber\\&-
{\nabla}^{\alpha }{ C _{\alpha }\,^{\beta\gamma}}\,   C 
_{\gamma}\,^{\delta}\,_{\beta}  w _{\delta}
+ C 
^{\alpha }  C 
_{\alpha 
}\,^{\beta\gamma}  C _{\gamma}\,^{\delta}\,_{\beta}  w _{\delta} -  C ^{\alpha 
\beta\gamma}  C _{\alpha \gamma}\,^{\delta}  C _{\delta}\,^{\mu}\,_{\beta}  w 
_{\mu}\big)
 + ( 4\,- 4 
\alpha_2) \big(
2{\nabla}^{\alpha }{ C 
^{\beta\gamma\delta}}\,   C _{\alpha \beta\delta}  w _{\gamma}
\nonumber\\& + 2 
{\nabla}^{\alpha 
}{ C ^{\beta\gamma\delta}}\,   C _{\beta \delta\alpha}  w _{\gamma} + 2 C 
^{\alpha 
\beta\gamma}  C _{\alpha 
}\,^{\delta}\,_{\gamma}  C _{\beta}\,^{\mu}\,_{\delta}  w _{\mu}  + w ^2  C 
^{\alpha 
\beta\gamma}  C _{\alpha  
\gamma\beta} 
 -  w ^2 C ^{\alpha \beta\gamma}  C 
_{\alpha 
\beta\gamma} 
 -\, { C ^{\beta\gamma\delta}}\,   C 
_{\beta \delta\gamma} 
 {\nabla}^{\alpha }{w _{\alpha } }\big)
\nonumber\\&
+ (16\, \beta_2- 8\, \beta_4- 4\, \alpha_2 -8 )  \big(C ^{\alpha \beta} 
{W}_{\alpha \beta}
-2 {\nabla}^{\alpha }{ C _{\alpha }\,^{\beta\gamma}}\,  {W}_{\beta\gamma}
+2 {W}^{\alpha \beta}  C ^{\gamma}  C _{\beta \gamma\alpha}
-2 {W}^{\alpha \beta}  C _{\alpha}\,^{\gamma}\,_{\beta}  w _{\gamma}
+ 2 {W}^{\alpha \beta}  C _{\alpha }  w _{\beta}
\big)
\nonumber\\&
+ ( 4 - 4\, \beta_2) \big(C ^{\alpha }  C ^{\beta}  w _{\alpha }  w _{\beta}
- w ^2  C ^2
+w^2  C ^{\alpha \beta\gamma}  C 
_{\gamma}\,^{\delta}\,_{\alpha }  
-2 {\nabla}^{\alpha }{C _{\alpha 
}\,^{\beta\gamma}}\,   C _{\gamma}  w _{\beta}
-2 {\nabla}^{\alpha }{ C ^{\beta}}\,   C _{\alpha }  w _{\beta} 
\nonumber\\&+ 2  C 
^{\alpha }  C _{\alpha }\,^{\beta\gamma}  C _{\beta\gamma}\,^{\delta}  w 
_{\delta}+ 2 
 C ^{\alpha }  C ^{\beta}  C _{\alpha }\,^{\gamma}\,_{\beta}  w _{\gamma} 
+2 w_\alpha C_\beta \nabla^\alpha C^\beta + 2  
C ^{\alpha }  C ^{\beta\gamma}\,_{\alpha }  C _{\beta}\,^{\delta}\,_{\gamma}  w 
_{\delta}
\big)
- 8\, 
{\nabla}^{\alpha }{ C 
_{\alpha }\,^{\beta\gamma}}\,  {\nabla}_{\beta}{ w _{\gamma}}\, 
\nonumber\\&+ 
8\, {\nabla}^{\alpha }{ w _{\gamma}} C _{\alpha }\,^{\gamma\beta}\,   w 
_{\beta}
 + (8 - 16\, \alpha_2 + 8\, \beta_2)  C ^{\alpha \beta\gamma}  C _{\alpha 
\gamma}\,^{\delta}  C 
_{\beta}\,^{\mu}\,_{\delta}  w _{\mu} 
+ 8\, {\nabla}^{\alpha 
}{ w ^{\beta}}\,   C _{\beta}\,^{\gamma\delta}  C _{\gamma \delta\alpha} - 8\, 
{\nabla}^{\alpha 
}{ w ^{\beta}}\,   C ^{\gamma}  C _{\alpha  \gamma\beta} 
\nonumber\\& 
+ 
(16 + 8\, \alpha_2 - 24\, 
\beta_2)  C ^{\alpha }  C _{\alpha }\,^{\beta\gamma}  w _{\beta} 
 w _{\gamma} + ( - 8 - 
8\, \alpha_2 + 16\, \beta_2)  C ^{\alpha \beta\gamma}  C _{\alpha 
\gamma}\,^{\delta} w _{\beta}  w _{\delta}  + 8\,\alpha_2 {\nabla}^{\alpha }{ C 
^{\beta\gamma\delta}}\,   C 
_{\alpha \beta 
\gamma}  w _{\delta} 
  \nonumber\\&   + ( - 8 + 4\, \alpha_2 + 4\, 
\beta_2)  C ^{\alpha \beta\gamma}  C _{\alpha }\,^{\delta}\,_{\gamma}  w 
_{\beta}  w _{\delta}+
(12 + 8\, \alpha_2 + 16\, \beta_4 - 32\, \beta_2) {W}^{\alpha \beta}  C 
_{\alpha  
\beta}\,^{\gamma}  w _{\gamma}  - 8\, \beta_2 
{\nabla}^{\alpha }{ C _{\alpha }\,^{\beta\gamma}}\,   C _{\beta}  w _{\gamma}
\nonumber\\&  
 + ( - 8\, \alpha_2 - 8 
+ 16\, \beta_2)  C ^{\alpha \beta\gamma}  C _{\alpha \gamma}\,^{\delta}  C 
_{\beta \delta}\,^{\mu} 
 w _{\mu} + ( - 8 + 16\, \beta_2) {\nabla}^{\alpha }{ C _{\alpha 
}\,^{\beta\gamma}}\,   
C _{\beta \gamma}\,^{\delta}  w _{\delta}
+ (8 - 16\, \beta_2) 
{\nabla}^{\alpha }{ C 
^{\beta}}\,   C _{\alpha \beta}\,^{\gamma}  w _{\gamma} 
 \nonumber\\
 &+ ( - 8\, \alpha_2 - 8 - 16\, \beta_4 + 32\, \beta_2) 
{W}^{\alpha \beta}  
C _{\alpha }\,^{\gamma\delta}  C _{\gamma \delta\beta} + ( - 4\, \alpha_2 - 4 + 
8\, \beta_2)  
C 
^{\alpha  \beta 
\gamma}  C _{\alpha \beta}\,^{\delta}  w _{\gamma}  w _{\delta}
\bigg].
\end{align}
\end{widetext}
It is worth mentioning that the term $\nabla_\alpha 
C^{\alpha\beta\gamma}W_{\beta\gamma}$ contains an interaction term between Weyl 
vector and torsion tensor which was assumed in \cite{isra}.

Finally, the full action of the theory can be written as
\begin{align}
S=\f{1}{2\kappa^2}\int 
d^4x\sqrt{-g}&\bigg[R-6w^2-C^2+C_{\alpha\mu\lambda}C^{\alpha\lambda\mu}
-4w^\alpha 
C_\alpha\bigg]+S_W+S_C+S_I.
\end{align}

Let us discuss about the coupling of the ordinary matter field to our theory. 
Note that we have 3 independent field in the theory. The metric tensor 
couples to the energy-momentum tensor of the matter and responsible for the 
gravitational force. The torsion tensor should be coupled to the spin tensor of 
the matter \cite{cartan}. Due to the works of Weyl to unify the electromagnetic 
field with gravity \cite{weyl}, one can imagine that the Weyl vector can be 
coupled to the electric current of the matter. In this sense, one can add the 
matter Lagrangian of the form
\begin{align}\label{}
S_m=\int 
d^4x\sqrt{-g}\mathcal{L}_m(\psi_i,g_{\mu\nu},T_{\rho\sigma\delta},w_\alpha),
\end{align}
where $\psi_i$ present the matter fields. By varying the matter Lagrangian wrt the metric one can obtain the 
energy-momentum tensor of the ordinary matter as
\begin{align}\label{}
T^{m}_{\mu\nu}=\f{-2}{\sqrt{-g}}\f{\delta(\sqrt{-g}\mathcal{L}_m)}{\delta 
g^{\mu\nu}}.
\end{align}
Variation of $\mathcal{L}_m$ wrt the torsion is the spin tensor
\begin{align}\label{}
\tau_{\mu}^{~\nu\rho}=2\f{\delta\mathcal{L}_m}{\delta T^{\mu}_{~\nu\rho}}.
\end{align}
In the special case where the matter field can be described by a scalar field, 
the spin tensor is zero and the torsion equation of motion is source-free. In 
other cases, one can define the spin tensor of the matter field from Noether's 
theorem and use it as a source of the torsion equation of motion.
The variation of $\mathcal{L}_m$ wrt the Weyl vector is
\begin{align}\label{}
J^{\mu}=2\f{\delta\mathcal{L}_m}{\delta w_{\mu}},
\end{align}
which can be considered as an electric current of ordinary matter. In the 
case of uncharged matter field the current $J^\mu$ will become zero and the 
Weyl equation of motion will become source-free. 

In the following we will 
assume that the matter Lagrangian vanishes for simplicity. The effect of matter 
fields and the interaction of torsion tensor and Weyl vector with it will be 
postponed to future works.

\section{Special Case for contortion tensor}\label{sec2}
The torsion tensor can be decomposed irreducibly into
\begin{align}\label{7}
T_{\mu\nu\rho}=\f 2 3 
(t_{\mu\nu\rho}-t_{\mu\rho\nu})+\f13(Q_{\nu}g_{\mu\rho}-Q_\rho 
g_{\mu\nu})+\epsilon_{\mu\nu\rho\sigma}S^\sigma,
\end{align}
where $Q_\mu$ and $S^\mu$ are two vector fields. The vector $Q_\mu$ is actually 
the trace of torsion over its first and third indices. The tensor 
$t_{\mu\nu\rho}$ is symmetric with respect to $\mu$ and $\nu$ and has the 
following
properties
\begin{align}\label{8}
t_{\mu\nu\rho}+t_{\nu\rho\mu}+t_{\rho\mu\nu}=0,\quad 
g_{\mu\nu}t^{\mu\nu\rho}=0=g_{\mu\rho}t^{\mu\nu\rho}.
\end{align}
 One can decompose the 
contortion tensor according to the above relation as
\begin{align}\label{9}
C_{\rho\mu\nu}=\f 4 3 
(t_{\mu\nu\rho}-t_{\rho\nu\mu})+\f23(Q_{\mu}g_{\nu\rho}-Q_\rho 
g_{\mu\nu})+\epsilon_{\rho\mu\nu\sigma}S^\sigma.
\end{align}

Let us assume that the contortion tensor has the following simple form
\begin{align}\label{9}
C_{\rho\mu\nu}=\hat{Q}_{\mu}g_{\nu\rho}-\hat{Q}_\rho g_{\mu\nu},
\end{align}
where $t_{\mu\nu\rho}=0$, $S^\sigma=0$ and 
$\hat{Q}_\mu=\frac{2}{3}Q_\mu$.

The action can then be expanded as
\begin{align}\label{16}
S&=\int d^4x\sqrt{-g}\bigg[\f{1}{2\kappa^2}\big(
R-6w^2-6\hat{Q}^2+12w^\alpha \hat{Q}_\alpha)
-4(1+\alpha_2-2\beta_2)\hat{Q}_{\mu\nu}\hat{Q}^{\mu\nu}
\nonumber\\&+8(2+\alpha_2+2\beta_4-4\beta_2)\hat{Q}_
{ \mu\nu } W^ { 
\mu\nu}-4(3+2\alpha_2+2\alpha_3-8\beta_2+16\beta_3+8\beta_4)W_{
\mu\nu } W^ {
\mu\nu }
\bigg].
\end{align}
In general, the above action may have some ghost and or tachyon instabilities. 
In order to 
examine this issue, we first diagonalize the kinetic and potential terms for 
$\hat Q_\mu$ 
and $w_\mu$ with the result
\begin{align}\label{52}
S=\int 
d^4x\sqrt{-g}&\bigg[\f{1}{2\kappa^2}R-\f{1}{4}X_{\mu\nu}X^{\mu\nu}-\f{1}{4}Y_{
\mu\nu}Y^{\mu\nu}
-\f {1}{2}m^2X_\mu X^\mu\bigg],
\end{align}
where $X_{\mu\nu}$ and $Y_{\mu\nu}$ are strength tensors respectively according 
to vectors $X_\mu$ and $Y_\mu$ which will be defined below. As one can see from 
the above action, the theory contains one massless and one 
massive vector fields with mass
\begin{align}\label{53}
m^2=\f{1}{2\kappa^2}\frac{3 (A+2 B+C)}{B^2-AC},
\end{align}
where we have defined
\begin{subequations}
\begin{align}\label{54}
A&=-4 - 4 \alpha_2 + 8 \beta_2,\\
B&=8 + 8 \beta_4 + 4 \alpha_2 - 16 \beta_2,\\
C&=-12 - 32 \beta_4 -64 \beta_3 - 8 \alpha_2 + 32 \beta_2 - 8 \alpha_3.
\end{align}
\end{subequations}
The new fileds can be related to the original fields $\hat Q_\mu$ and $w_\mu$ as
\begin{align}
&X_\mu=\f{2}{\sqrt{2\beta^2-1}}\bigg[(\alpha\beta\lambda_{+}-\lambda_{-}\sqrt{
(1-\alpha^2)(\beta^2-1)})\hat Q_\mu+(\beta\lambda_{+}\sqrt{
1-\alpha^2 } +\alpha\lambda_ {-}\sqrt{\beta^2-1})w_\mu\bigg], \label{55}\\
&Y_\mu=-\f{2}{\sqrt{2\beta^2-1}}\bigg[(\beta\lambda_{-}\sqrt{1-\alpha^2}
+\alpha\lambda_ {+}\sqrt{\beta^2-1})\hat{Q}_\mu-(\alpha\beta\lambda_{-}-\lambda_{+}\sqrt{
(1-\alpha^2)(\beta^2-1)})w_\mu\bigg],\label{552}
\end{align}
where
\begin{align}\label{56}
&\alpha=\left[\f{1}{2}\left(1+\f{A-C}{\sqrt{4B^2+(A-C)^2}}\right)\right]^{\f{1}{
2}} , \\
&\beta=-\left[\f{1}{2}\left(\f{(A+2B+C)\sqrt{4B^2+(A-C)^2}}{4B^2+(A-C)^2+2B(A
+C)} +1\right)\right]^{\f{1}{2}},
\end{align}
and
\begin{align}\label{57}
\lambda^2_{\pm}=-\f{1}{2}\left(A+C\pm\sqrt{4B^2+(A-C)^2}\right).
\end{align}
In order to have a ghost and tachyon free theory we should have $m^2>0$ and the 
new fields \eqref{55} and \eqref{552} should be meaningful. We thus conclude 
that the 
parameters $\alpha_2,\alpha_3,\beta_2$ and $\beta_4$ should satisfy the 
relations
\begin{align}\label{58}
m^2>0, \qquad\lambda^2_{\pm}>0,
\end{align}
together with reality of square roots. 

The action \eqref{52} can be seen as a gravitational theory minimally coupled 
to a 
Proca field and a Maxwell field, i.e. the Einstein-Maxwell-Proca theory. 
Geometrization of the Einstein-Proca (EP) action was done before with different 
approaches. In \cite{tomi}, the EP action was obtained by writing the 
Gauss-Bonnet action in the Weyl space-time. However, in the Weyl 
space-time, adding the contraction of the antisymmetric part of the Ricci 
tensor with the Ricci tensor itself, is sufficient to produce the EP action, 
which was done in \cite{buch}. One can also obtain the EP action by writing the 
$f(R,R_{\mu\nu}R^{\mu\nu})$ action in the Palatini formalism \cite{soti}. In 
this case the equation of motion for the connection can be solved for the 
connection itself, leading to an expression which is equivalent to the Weyl 
connection with the Weyl vector $w_\mu=\f{1}{f^\prime}\partial_\mu f^\prime$ where prime is taking derivative with respect to the function's argument. 
The connection of this theory and the resulting Ricci tensor is asymmetric. The 
antisymmetric part of the Ricci tensor can then produce a kinetic term for the 
Weyl vector $w_\mu$.

In this paper, we have 
used the Weyl-Cartan space-time for writing the Gauss-Bonnet action. In the 
special case where only the trace part of the torsion tensor is non-vanishing, 
the theory reduces to a bi-vector-tensor theory, which can not be obtained in 
the Weyl space-time. Note that writing the Gauss-Bonnet action in the Cartan 
space-time and keeping only the trace part of the torsion will produce the EP 
action \footnote{This can be seen from equation \eqref{16} by assuming 
$w_\mu=0$}. 

The theory presented here can be seen as a gravitational theory 
coupled to two vector galileon fields \cite{vecgali}. Because we have added 
only a Gauss-Bonnet term (which is the second Lovelock term) to the action, 
only 
the second vector galileon term appears. One may expect that adding higher 
order Lovelock terms to the action may produce higher order vector galileon 
terms \footnote{One may also expect that the $S^\mu$ term in the decomposition 
of torsion can produce vector galileon term because of the appearance of 
Levi-Civita tensor. This possibility will be considered in a separate work.}.
One should note that in the vector galileon theories, the helicity-0 part plays 
the role of galileon fields \cite{gali}. Galileons are some scalar fields which 
has higher than second order time derivatives in the action, but produce at 
most 
second order field equations. The resulting action is known as the Horndeski 
action \cite{horn}. Note that in the action \eqref{52} the helicity-0 part 
vanishes due to the special form of the kinetic term. However, considering the 
theory in higher dimensions and then performing the Kaluza-Klein reduction, one 
can obtain the Horndeski theory \cite{tomi}.

Our theory can be seen as a generalization of the Horndeski 
action, in the sense that we have also added some higher order curvature tensor 
terms to the action. However, we have used the Gauss-Bonnet combination 
which causes that the action has no higher than second order time derivatives.

Conditions \eqref{58} 
can be solved analytically in terms of $A,B$ and $C$. The solution implies that 
$C<0$ and $A\in (C,0)$. The parameter space is divided into 
three parts as follows
\begin{align}\label{}
&C<A<\f13 C \Rightarrow\begin{cases}
0<B<-\f14\bigg(A+C+\sqrt{(3C-A)(3A-C)}\bigg)\\\textrm{or}
\\-\f14\bigg(A+C-\sqrt {
 (3C-A)(3A-C)}\bigg)<B<\sqrt{AC}
 \end{cases}\nonumber\\
&A=\f13 C\quad\Rightarrow\quad 0<B<-\f{\sqrt{3}}{3}C,\quad B\neq
-\f13 C,\nonumber\\
&\f13 C<A<0\quad\Rightarrow\quad 0<B<\sqrt{AC}.
\end{align}
Solving the above conditions in terms of $\alpha_i$ and $\beta_i$ is very 
difficult to obtain. So, in the following we will study some special cases.
\subsection{Case I: $\beta_3=\beta_2=\alpha_3=0$ and $\alpha_2=1$}
In this case the the action $S_G$ reduces to
\begin{align}\label{58.1}
S_G=&\int d^4x\sqrt{-g}\bigg[K^{\alpha \beta 
\gamma\delta} 
K_{\gamma \delta \alpha \beta}
- 4 \mathcal{K}_{\beta\gamma} {\mathcal{K}}^{\beta\gamma}
-4\beta_4
K_{\alpha \beta} \mathcal{K}^{\alpha\beta}
+ K^2\bigg],
\end{align}
and the constraints \eqref{58} satisfy if
\begin{align}\label{59}
-\f12(1-\sqrt{2})<\beta_4<\f12(1+\sqrt{2}).
\end{align}
\subsection{Case II: $\beta_2=0=\alpha_3$ and $\beta_3=0=\beta_4$}
In this case the the action $S_G$ reduces to
\begin{align}\label{59.1}
S_G=&\int d^4x\sqrt{-g}\bigg[(1-\alpha_2) K^{\alpha \beta 
\gamma\delta} K_{\alpha \beta \gamma\delta} + \alpha_2 K^{\alpha 
\beta 
\gamma\delta} 
K_{\gamma \delta \alpha \beta}
- 4{\mathcal{K}}_{\beta\gamma} {\mathcal{K}}^{\beta\gamma}
+ K^2\bigg],
\end{align}
and the constraints \eqref{58} satisfy if
\begin{align}\label{60}
\frac{2}{1+\sqrt{5}}<\alpha_2<2.
\end{align}
\subsection{Case III: $\beta_2=0=\alpha_3$ and $\beta_3=-\beta_4$.}
In this case the the action $S_G$ reduces to
\begin{align}\label{59.1}
S_G=&\int d^4x\sqrt{-g}\bigg[(1-\alpha_2) K^{\alpha \beta 
\gamma\delta} K_{\alpha \beta \gamma\delta}+ \alpha_2 K^{\alpha 
\beta 
\gamma\delta} 
K_{\gamma \delta \alpha \beta}
- 4 {\mathcal{K}}_{\beta\gamma} {\mathcal{K}}^{\beta\gamma} 
+4 \beta_4 (K_{\alpha \beta} {K}^{\alpha \beta}-
K_{\alpha \beta} {\mathcal{K}}^{\alpha\beta})
+ K^2\bigg],
\end{align}

\begin{center}
 \includegraphics[scale=0.3]{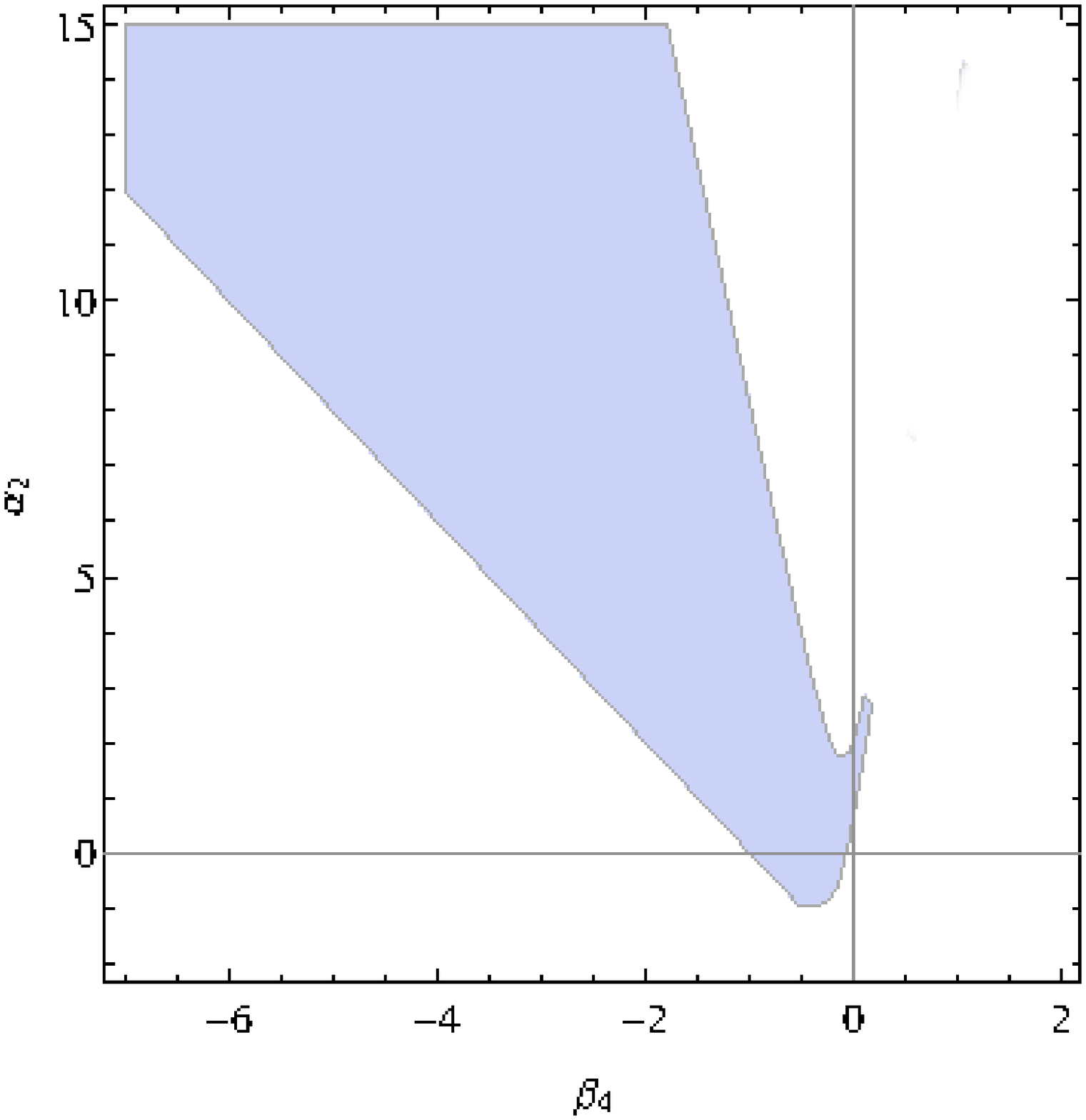}\\
 \small{FIG1: Allowed range of $\beta_4$ and $\alpha_2$ for the action 
\eqref{59.1} to 
become ghost and tachyon free theory.}
 \label{fig12}
\end{center}
In the figure we have plotted the allowed region of parameter space 
($\alpha_2$,$\beta_4$) in order to have a ghost and tachyon free bi-vector 
theory.
\section{conclusion}
In this paper, we have introduced a ghost free modified theory of 
gravity by generalizing the geometry to be the Weyl-Cartan space-time. Using 
the standard Einstein-Hilbert term for this geometry, the action reduces to the 
Ricci scalar, plus possible mass terms for Weyl vector and the torsion tensor. 
In this case, kinetic terms for these two new fields are not produced. In 
order 
to make the Weyl vector and the torsion tensor dynamical, one can add some 
kinetic terms by hand, which was done in \cite{zahra}. 

In this paper, in order 
to produce kinetic terms for the Weyl and torsion fields, we have generalized 
the action to be of Gauss-Bonnet type. In 4D Riemannian geometry the 
Gauss-Bonnet term becomes a 
total derivative and dropped from the action. However in the Weyl-Cartan 
geometry, this term produces a bunch of interaction and kinetic terms for Weyl 
and 
torsion fields. In the Weyl-Cartan geometry the curvature tensor has less 
symmetries than the Riemann tensor. So one can write more than three quadratic 
terms according to the curvature. In general, the resulting action does not 
reduce to the 
standard Gauss-Bonnet action. However, demanding that the action should reduce 
to the standard Gauss-Bonnet action in the case of vanishing Weyl and Cartan, 
all the higher than two time derivatives of the action becomes a total 
derivative and do not contribute to the action.

For further considerations, we have studied a special case of the theory where 
only the 
trace part of the torsion tensor is non-zero. In this case, the theory is 
reduced to general relativity plus one massive and one massless vector fields. 
This theory can be seen as an Einstein-Maxwell-Proca system which has 7 d.o.f.,
two for the graviton, two of the Maxwell field and three for the Proca vector 
field. The Einstein-Proca system was also geometrized in the context of Weyl 
gravity in \cite{tomi,buch} and also in the context of Palatini-f(R) theories 
\cite{soti}. However, our theory in its reduced form where we have considered 
in section \ref{sec2}, has two vector fields minimally coupled to gravity. One 
can see that the trace of the torsion tensor can also play the role of a Proca 
field. 

It will be very interesting to note that, in order to obtain a kinetic 
term for a vector field by adding higher order curvature terms to the action, 
the Ricci tensor should be asymmetric. In Weyl space-time considered in 
\cite{tomi,buch} and also in the Weyl-Cartan space-time considered here, the 
Ricci tensor is asymmetric. In \cite{soti}, the Palatini assumption is 
sufficient to take the connection asymmetric, and as a result the Ricci tensor 
will be asymmetric.

In this paper we showed that the absence of Ostrogradski ghost reduces the 
dimension of parameter space of the theory to 5. However, this will not prove 
that the theory is healthy. The full theory may has some Boulware-Deser ghosts 
or tachyonic instabilities. In section \ref{sec2} we have considered some 
special cases of the theory and obtained the healthy region of the parameter 
space of the theory. 

It is worth mentioning that higher order Lovelock terms can also contribute to the action in 4D in the context of Weyl-Cartan theory. In this paper we have only studied the second order term to show that the formalism is capable of producing kinetic terms for Weyl vector and torsion tensor. Higher order terms can be responsible for higher order self-interaction terms for the Weyl and torsion which may resembles vector Galileons and its generalizations. This issue should be considered carefully in more details which should be studied in a separate work.

A question which should be answered in future works is about physics of our 
model. This can be done by solving for both cosmological and static solutions 
at the level of background to see if self-accelerating solutions exists and the solar system tests are satisfied. The other issue is coupling to matter which definitely 
affects predictions of our model which is beyond the scope of current work.

\end{document}